\documentclass[aps,pre,preprint,superscriptaddress,showpacs]{revtex4-1}
\usepackage{graphicx}
\usepackage{amssymb}
\usepackage{epsfig}
\usepackage{amsmath}
\usepackage{times}
\usepackage{bm}
\usepackage{color}
\usepackage{array}
\usepackage{mathrsfs}
\usepackage{subeqnarray}
\usepackage{cases}
\usepackage{float}

\begin{document}

\title{Agent-based simulations of China inbound tourism network}

\author{Jinfeng Wu}
\affiliation{School of Geography and Tourism, Shaanxi Normal University, Xi'an 710062, China}

\author{Xingang Wang}
\email{E-mail: wangxg@snnu.edu.cn}
\affiliation{School of Physics and Information Technology, Shaanxi Normal University, Xi'an 710062, China}

\author{Bing Pan}
\affiliation{Department of Recreation, Park, and Tourism Management, School of Health and Human Development, Pennsylvania State University, University Park, Pennsylvania 16802, USA}

\begin{abstract}

Based on the results of a large-scale survey, we construct an agent-based network model for the independent inbound tourism of China and, by the approach of numerical simulation, investigate the dynamical responses of the tourist flows to external perturbations in different scenarios, including the closure of a tourist city, the opening of a new port in western China, and the increase of the tourism attractiveness of a specific city. Numerical results show that: (1) the closure of a single city in general will affect the tourist visitations of many other cities in the network and, comparing to the non-port cities, the overall visitation volume of the system is more influenced by closing a port city; (2) the opening of a new port city in western China will attract more tourists to the western cities, but has a negligible impact on either the overall visitation volume or the imbalanced tourist distribution; and (3) the increase of the tourism attractiveness of a non-port (port) city normally increases (decreases) the overall visitation volume, yet there are exceptions due to the spillover effect. Furthermore, by increasing the tourism attractiveness of a few cities simultaneously, we investigate also the strategy of multiple-city-upgrade in tourism development. Numerical results show that the overall tourist volume is better improved by upgrading important non-port cities that are geographically distant from each other. The study reveals the rich dynamics inherent in complex tourism network, and the findings could be helpful to the development and management of China inbound tourism. 

\end{abstract}

\date{\today}
\maketitle

\section*{Introduction} 

A well-known feature of the tourism system is that cities inside one country form very complicated collaborative-competitive relationships~\cite{LN:book,NS:book}. On one hand, a tourist city acts not only as a destination, but also the transition stops for the tourists to transfer to other cities. In this regard, the increase of the visitation of one city will lead to the increase of the visitation of the transition city, i.e., their visitations will be positively correlated and they form a collaborative relationship. On the other hand, due to time limit and travel cost, a tourist can only visit a limited number of cities in the destination country. Therefore, the increase of tourist visitation of a city could decrease the visitation of another one. In such a case, the visitations of these cities could be negatively correlated and they form a competitive relationship. The collaborative-competitive relationships between tourist cities render the dynamics of tourism system highly nonlinear, giving rise to many intriguing collective behaviors, e.g., a small and local perturbation on a single city might trigger a large and global event over the whole system. The complex relationships and nonlinear dynamics pose unprecedented challenges for the development and management of modern tourism system, thus calling for the development of new theoretical models and analysis methods~\cite{HYH:2006,BR:2011,WY:2007,RMR:2016}. 

As the $4$th largest inbound tourist destination in the world, the inbound tourism is an essential component of the China's tourism industry~\cite{WTO}. However, China has been the largest origin country of outbound tourists worldwide since 2009. Thus, comparing with the outbound tourism, the development of China's inbound tourism is largely lagged behind~\cite{CTA}. Moreover, among the inbound tourists, a large fraction (about $81\%$) is contributed by oversea Chinese tourists from Hong Kong, Macao, and Taiwan~\cite{CTA}. The dramatic imbalance between the inbound and outbound tourism markets makes the development of China's inbound tourism an economic imperative due to trade deficit, and how to eliminate this deficit has been an active topic for both the policy-makers and researchers in the past years~\cite{LF:2010,WJF:2010,WJF:2014,MY:2014,YX:2013}. Besides the overall tourist volume, another key challenge in developing the China's inbound tourism is how to deal with the problem of imbalanced tourist distribution. According to the recent reports~\cite{Reports}, the tourism resources of China are concentrated in cities in the eastern area. These eastern cities, which are also densely populated and well developed in economy, attracted most of the inbound tourists (more than $80\%$)~\cite{YX:2014}. Despite economic stimulations such as infrastructure investments and policy supports, the western cities are still lagging behind in terms of tourist visitation volume~\cite{ZH:2008}. The imbalanced distribution harms the performance of the whole inbound tourism, and how to reduce the imbalance by attracting more tourists to western cities has been another key issue for the development of a healthy inbound tourism industry. To solve the problem of imbalanced tourist distribution, a common approach currently adopted in practice is to increase the tourism attractiveness of some western cities~\cite{ZH:2008}, e.g., improving their service quality and transportation infrastructure. However, due to the complex collaborative-competitive relationships between the tourist cities, the upgrading of a city may either increase or reduce the overall visitations of the China's inbound market. To have a proper evaluation on the impact of city upgrading, it is necessary to consider the networked tourist cities globally from the perspective of a dynamic and complex system. 

Network science provides a powerful tool for exploring the structure and performance of complex tourism systems~\cite{NS:book,RB:2010}. Stimulated by the discovery of the small-world and scale-free features in many real-world networks~\cite{SW-model,SFN-model}, in the past two decades, researchers have spent great efforts on the exploration of the topological properties of various empirical networks, as well as their influences on the network functions~\cite{SB:2006}. As a typical example of complex social network, a variety of tourism network models have been constructed and studied in the past years~\cite{WJF:2002,NS:2008,CNG:2014,CC:2016,BL:2017}. In the general model of tourism network, the destinations are represented by nodes, and transportations among destinations are represented by links. Depending on the specific problem that is interested, the network nodes and links may have different definitions. For instance, treating different tourism sectors within a destination city (e.g., the attractions, hotels, economic stakeholders, and service providers) as nodes and the business relationship between them as links, in Refs.~\cite{BR:2013,RB:2010,NS:2010,NS:2008,CNG:2014,PHS:2016,RB:2016} the authors have constructed micro-tourism network models. At the macroscopic level, a node can be defined as a country or a group of cities inside one area, and links can be defined as the economic ties among the countries or areas~\cite{WJF:2002,BL:2017,MJIL:2008,HT:2015}. In previous studies of complex tourism networks, a common finding is that the networks possess the general features of many complex networks, e.g., the small-world property and heterogeneous degree distribution~\cite{MJIL:2008,CNG:2014}. It is worth noting that the existing studies focus mainly on the topological properties of tourism networks, whereas their dynamic properties have been largely overlooked~\cite{WJF:2002,NS:2008,CNG:2014,CC:2016,BL:2017,BR:2013,RB:2010,NS:2010,NS:2008,CNG:2014,PHS:2016,RB:2016,WJF:2002,BL:2017,MJIL:2008,HT:2015}. Specifically, it remains unclear how the patterns of tourist flows emerge from the random tourist motions, and to what extend the tourism market will be influenced if a destination city is closed during a tourism crisis~\cite{PHS:2016,HT:2015,SHY:2006,LA:2006,LXY:2012,YY:2013,KP:2014,GYR2015,AV:2016}. These dynamic properties, which are rooted in the collaborative-competitive relationships between the tourist cities, are crucially important for the stable functioning of a tourism network, as well as for the development and management of modern tourism.

In the present work, we construct an agent-based network model for China's inbound tourism system~\cite{YX:2013,YX:2014}, and, with numerical simulations, explore its dynamical responses to various external perturbations. In the constructed network model, each node stands for a specific tourist city in China, and two nodes are connected by a link if there is a direct transportation between the corresponding cities. We introduce a large number of agents into the network through the port cities, and let the agents travel among the connected nodes in a random fashion. The results of a large-scale survey provide the key parameter characterizing the random movement of the agents. Based on the constructed model, we then investigate the detailed responses of the tourist flows under different scenarios of practical interest, including: (1) the removal of a city from the network (which simulates the closure of a city due to tourism crisis); (2) the opening of a new port city in western China (which models the adjustment of the tourism policy in order of solving the imbalanced tourist distribution); and (3) the increase of city tourism attractiveness (which corresponds to the city upgrading). Our main finding is that in a complex tourism network, the performance of the cities are strongly coupled with each other, and the change of a single city could affect the tourist volumes of many other cities, resulting in a large variation in the overall tourism market. The study sheds lights on the dynamics of tourist flows in complex tourism system, and could be helpful to the development and management of China's inbound tourism.

\section*{Results}
\noindent\paragraph*{\bf Conceptual framework for the agent-based tourism network.} The agent-based network of China's inbound tourism is constructed as follows. Firstly, we regard each destination city in China as a node. The cities are divided into two groups: the port and non-port cities. For the port cities, the tourists are allowed to enter or leave China according to certain probabilities; for the non-port cities, the tourists can only pass through them to another city. Correspondingly, the network nodes are classified into the port and non-port nodes. Secondly, we attribute each city (node) a weight, $a_i$, which characterizes the overall tourism attractiveness of the city (see {\bf Methods} for the calculation of $a_i$). Thirdly, a link is established between nodes $i$ and $j$ if the two cities are connected by either the direct flight or railway transportation. The links are weighted by the geographical distance, $d_{ij}$, between cities $i$ and $j$. Finally, we input a large number of agents, $N$, into the network through the port nodes, and let the agents travel among the nodes in a random fashion. To be specific, denoting $V_i=\{j\}$ as the set of neighboring nodes of $i$, the probability for an agent to be moving from node $i$ to $j$ is defined as (the gravity model)~\cite{PRL:2008,BF:2016}
\begin{equation}
p_{ij}=\frac{a_j^\gamma/d_{ij}}{\sum_{j\in V_{i}}(a_j^\gamma/d_{ij})},
\label{eq1}
\end{equation}
with $\gamma$ a key parameter to be estimated from the surveyed results.

By numerical simulations, we monitor the movement of the agents on the network and record their travel routes. When an agent arrives a port city, there is a predefined probability for it to leave the network. If the agent departs from this port city, the number of agents in the network will be deceased by one. The simulation is stopped once all agents have left the network. We then conduct analysis on the statistical properties of the recorded travel routes (itineraries). We characterize each travel route by a chain of nodes, which are always started from and ended at the port nodes. For example, if an agent visited four cities in sequence, we denote its travel route by the chain: $A\rightarrow B\rightarrow C\rightarrow D$, with $A$ the port city from where the agent enters the network, $D$ the port city from where the agent leave the network, and $B$ and $C$ the transition cities the agent passes through. The length of the travel route is defined as $L=m+1$, with $m$ the number of transition cities. For the example mentioned, we have $L=3$. The first statistical property we are interested is the averaged route length, $\left<L\right>=\sum_{l=1}^N{L_l}/N$. For the fixed number of agents, the larger is $\left<L\right>$, the larger will be the overall visitation volume of the network. So the value of $\left<L\right>$ reflects in fact the overall performance of the system. The second statistical property we are interested is the distribution of the tourist flows. Here, tourist flow is defined as the total amount of tourists passing through a specific link. For the link between nodes $i$ and $j$, the tourist flow is denoted by $w_{ij}$. It is noted that as $w_{ij}$ includes both the routes from $i$ to $j$ and from $j$ to $i$, we therefore have $w_{ij}=w_{ji}$. The third statistical property we are interested is the visitation volumes of the individual cities, $s_i=\sum_{j=V_i}w_{ij}$, i.e., the total number of agents visiting node $i$. Different from $\left<L\right>$, the matrices $\{w_{ij}\}$ and $\{s_i\}$ capture the collective behaviors of the agents at the link and nodal levels, and represent the detailed characteristics of the tourist distribution over the network. It is worth noting that for the fixed number of agents, the three statistical properties are related with each other: $N\left<L\right>=\sum_i s_i=\sum_{i>j} w_{ij}$.      

\paragraph*{\bf Empirical and surveyed results.} We concrete the conceptual model of tourism network by the results of a large-scale survey. In constructing the network model, we need to estimate the following key parameters. Firstly, we need to identify the set of cities to be contained in the network. Currently there are totally about $700$ tourist cities in China, but not all of them are captured in our surveys. Secondly, as the behaviors of the tourists are very different at the port and non-port cities, we need to identify the set of port cities. Thirdly, we need to estimate the fraction of tourists arriving a specific port city, as well as the probability for a tourist to depart from it. These parameters can only be obtained from questionnaires. With these concerns, we asked the surveyed tourists to provide the complete information of their itineraries, including the arrival, departure and transition cities (see {\bf Supplementary} for an example of the completed questionnaires).  

The questionnaire was translated into five different languages (English, Japanese, Korean, Russian, and French), and administered from November 2010 to August 2011. The tourists were intercepted and surveyed at the major attractions of eight popular tourist cities in China, including Guangzhou, Shanghai, Hangzhou, Suzhou, Beijing, Chengdu, Guilin, and Xi'an. A total number of $3,000$ questionnaires were distributed, with $2,687$ returned on-site for a response rate of $89.6\%$. As we are focusing on only the behavior of independent foreign tourists, we extract the surveys from tourists who arranged their trips by themselves, or by their relatives, employers or other individuals, rather than by tourism companies or travel agencies. In addition, we remove the surveys collected from the overseas Chinese visitors (tourists from Hong Kong, Macao, and Taiwan), as they behave very differently from the foreign visitors~\cite{LW:2007}. In the end, we have $1,451$ effective questionnaires in total. Among them, $856$ itineraries contain at least two cities, i.e., with the route length $L\ge 1$. These itineraries are the final samples used in our model construction. An analysis of the final examples shows that the tourists came from $61$ different countries and areas: $37.7\%$ from Europe, $27.9\%$ from Asian countries, $14.8\%$ from Africa, $6.5\%$ from North America, $8.2\%$ from South America, and $4.9\%$ from Oceania. 

The itineraries contain a total of $n=58$ tourist cities in China, including $4$ port cities and $54$ non-port cities. The four port cities are Beijing ($i=1$), Shanghai ($i=2$), Guangzhou ($i=3$), and Hong Kong ($i=4$), and their arrival fractions of tourists are $p^a_1=49.0\%$, $p^a_2=26.4\%$, $p^a_3=13.2\%$ and $p^a_4=11.5\%$, respectively. The probability for a tourist to depart from a port city is defined as the ratio between the number of departure tourists and the total tourist visitations of the port city. In calculating the total tourist visitations of a port city, the number of arrival tourists has been excluded. From the surveyed results, we have the following departure probabilities: $p^d_1=72.7\%$ for Beijing, $p^d_2=57\%$ for Shanghai, $p^d_3=48.6\%$ for Guangzhou, and $p^d_4=51.1\%$ for Hong Kong. In Tab. \ref{table1}, we list all $58$ tourist cities by the descending order of their tourist visitations ($s_i$), together with their tourism attractiveness ($a_i$).

\begin{table}[htbp]
\centering
\setlength{\abovecaptionskip}{0pt}%
\setlength{\belowcaptionskip}{10pt}
\caption{The tourist cities appeared in the questionnaires. The cities are indexed by the descending order of their tourist visitation volumes, $s_i$. Tourism attractiveness, $a_i$, characterizes the overall attraction of a city. The four port cities are Beijing, Shanghai, Guangzhou, and Hong Kong.}
\begin{tabular}{p{1cm}<{\centering} p{2cm}<{\centering} p{2cm}<{\centering} p{2cm}<{\centering} p{1cm}<{\centering} p{1cm}<{\centering} p{2cm}<{\centering} p{2cm}<{\centering} p{2cm}<{\centering}}
\toprule
Index & City & Tourism attractiveness & Visitation volume & \vline & Index & City & Tourism attractiveness & Visitation volume\\
\hline
1   & Beijing & 224 & 970 & \vline&30 & Jiuzhaigou & 36 & 14\\
2   & Shanghai & 141 & 864 & \vline&31 & Wenzhou & 35 & 14\\
3   & Guangzhou & 76 & 529 & \vline&32 & Zhangjiajie & 29 & 14\\
4   & Hong Kong & 62 & 492 & \vline&33 & Guiyang & 27 & 12\\
5   & Xi'an & 66 & 467 & \vline&34 & Harbin & 47 & 11\\
6   & Guilin & 96 & 446 & \vline&35 & Luoyang & 50 & 10\\
7   & Hangzhou & 90 & 331 & \vline&36 & Shenyang & 41 & 10\\
8   & Chengdu & 80 & 241 & \vline&37 & Xining & 23 & 10\\
9   & Suzhou & 102 & 220 & \vline&38 & Yantai & 52 & 10\\
10   & Kunming & 32 & 131 & \vline&39 & Changchun & 16 & 8\\
11   & Chongqing & 180 & 110 & \vline&40 & Changsha & 52 & 8\\
12   & Nanjing & 49 & 76 & \vline&41 & Zhuhai & 6 & 7\\
13   & Tianjin & 82 & 71 & \vline&42 & Changzhou & 34 & 6\\
14   & Shenzhen & 28 & 65 & \vline&43 & Fuzhou & 30 & 6\\
15   & Xiamen & 38 & 53 & \vline&44 & Qinhuangdao & 50 & 6\\
16   & Dali & 17 & 40 & \vline&45 & Dunhuang & 18 & 4\\
17   & Dalian & 46 & 40 & \vline&46 & Hefei & 54 & 4\\
18   & Lijiang & 28 & 38 & \vline&47 & Jingdezhen & 20 & 4\\
19   & Wuhan & 58 & 36 & \vline&48 & Lhasa & 16 & 4\\
20   & Yiwu & 47 & 36 & \vline&49 & Lanzhou & 12 & 4\\
21   & Datong & 8 & 27 & \vline&50 & Nanchang & 18 & 4\\
22   & Huangshan & 76 & 26 & \vline&51 & Nanning & 36 & 4\\
23   & Wuxi & 75 & 26 & \vline&52 & Zhengzhou & 32 & 4\\
24   & Qingdao & 65 & 24 & \vline&53 & Haikou & 9 & 2\\
25   & Urumqi & 35 & 22 & \vline&54 & Xishuangbanna & 29 & 4\\
26   & Ningbo & 86 & 18 & \vline&55 & Nantong & 23 & 2\\
27   & Shangri-La & 20 & 16 & \vline&56 & Shantou & 15 & 2\\
28   & Sanya & 38 & 15 & \vline&57 & Xuzhou & 30 & 2\\
29   & Jinan & 38 & 14 & \vline&58 & Yichang & 51 & 2\\
\hline
\end{tabular}
\label{table1}
\end{table}

We next define the links connecting the nodes. A link is established between two nodes if there is at least one direct flight or direct railway connection between the corresponding cities. The flight information and flight distance between the connected cities are obtained from the Civil Aviation Administration of China (CAAC)~\cite{CAAC}. The railway information is obtained from the Service Center of National Railway Administration of China~\cite{NRA}. By the time the surveys were conducted, there are totally $326$ flight and railway connections among the investigated tourist cities. The connectivity of the cities is characterized by the distance matrix $\mathbf{D}={d_{ij}}$, with $d_{ij}=d_{ji}$ the flight or railway distance between cities $i$ and $j$.  

\begin{figure*}
\centering
\includegraphics[width=0.8\linewidth]{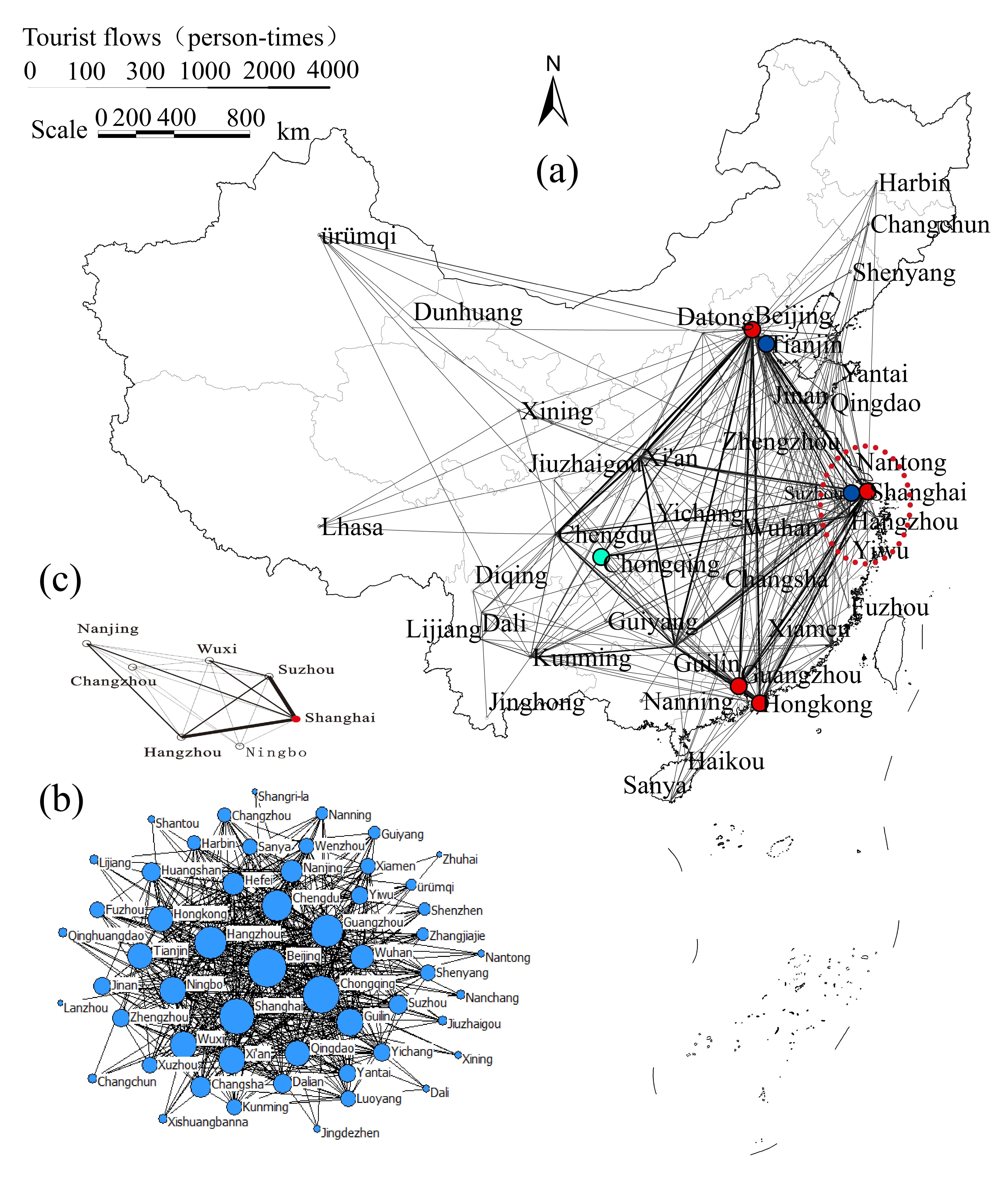}
\caption{{\bf The China inbound tourism network constructed from the empirical and surveyed data.} (a) The network structure, which consists of $n=58$ nodes (cities) and $326$ links of flight and railway transportations. The thickness of the link is proportional to the tourist flow. The four port cities are colored in red. Chongqing is colored in green. Tianjin and Suzhou are colored in blue. Dashed ellipse denotes the Yangtze-River-Delta area. (b) The heterogeneous distribution of the city visitation volumes, $\{s_i\}$. The size of node $i$ is proportional to its visitation volume, $s_i$. (c) The subnetwork formed by the cities inside the Yangtze-River-Delta area.}
\label{fig1}
\end{figure*}

By the empirical and surveyed results, we plot in Fig. \ref{fig1}(a) the tourism network embedded in the geographical space. In Fig. \ref{fig1}(a), the volume of the tourist flow between cities $i$ and $j$, $w_{ij}$, is represented by the thickness of the associated link. We see that: (1) in consistent with the results in previous studies~\cite{MJIL:2008,CNG:2014}, the tourism network possesses complex topological structure; (2) the distribution of the tourists is heterogeneous and imbalanced (most of the tourists are concentrated in the eastern part of China, especially in the well-developed regions such as the Yangtze-River-Delta and Pearl-River-Delta areas)~\cite{YX:2014}; (3) there are a few hub cities, e.g., the four port cities, which are densely connected to other cities in the network; (4) the tourist flows between the hub cities are clearly larger than those between the non-hub cities, forming the backbone of the network. The distribution of the tourist volumes, $s_i$, are plotted in Fig. \ref{fig1}(b) by the core-periphery fashion, i.e., cities of larger visitation volumes are arranged at the core and those of small visitation volumes at the periphery. We see that, besides the four port cities (Beijing, Shanghai, Guangzhou, and Hong Kong), the visitation volumes of Hangzhou, Chengdu, Suzhou, Xi'an, and Guilin are also very large, signifying their important roles in the system.  

\paragraph*{\bf Estimating the dynamic parameter of agent movement.} By far, we have fixed only the properties of the network nodes and links, including the sets of port and non-port nodes, the arrival percentages ($p^a_{1,2,3,4}$) and departure probabilities ($p^d_{1,2,3,4}$) of the port cities, the tourism attractiveness of each node ($a_i$), and the weights of the links (i.e., the distance matrix $\{d_{ij}\}$). We proceed to characterize the dynamic feature of the network, i.e., the moving fashion of the agents. As the node attractiveness $\{a_i\}$ and link weights $\{d_{ij}\}$ have been fixed, the random movement of the agents, as described by Eq. (\ref{eq1}), is dependent of the exponent $\gamma$ only. The parameter $\gamma$ is trained by comparing the surveyed results with the numerical results, as follows. We scan $\gamma$ over a wide range and, for each value of $\gamma$, compare the numerical results with the surveyed results for the following statistical distributions: the route length ($\{L_l\}$), the tourist flows ($\{w_{ij}\}$) , and the visitation volumes ($\{s_i\}$). For each distribution, the optimal value of $\gamma$ is defined as the one that minimizes the difference between the numerical and surveyed results. We thus have three different optimal values for the exponent $\gamma$. The final value of $\gamma$ used in Eq. (\ref{eq1}) is chosen as the average of the three optimal values, which is $\gamma\approx 2.9$ (see {\bf Methods} for more details on the estimation of $\gamma$).  

\paragraph*{\bf Numerical results.} With the constructed network, we next investigate numerically the collective behaviors of the tourists in different scenarios. The specific questions to be addressed are: (1) to what extent will the overall tourist volume be affected when a single city is closed? (2) will the opening of a new port city in western China attract more tourists from the eastern cities? and (3) how to improve the overall tourist volume by increasing the tourism attractiveness of one or a few cities? These questions capture the essential features of the collaborative-competitive relationships between cities in the tourism network, and also provide direct guidance to the development and management of the inbound tourism of China~\cite{LN:book,NS:book,HYH:2006,BR:2011,WY:2007,RMR:2016,PHS:2016,HT:2015,SHY:2006,LA:2006,LXY:2012,YY:2013,KP:2014,GYR2015,AV:2016}.

\begin{figure}
\centering
\includegraphics[width=0.7\linewidth]{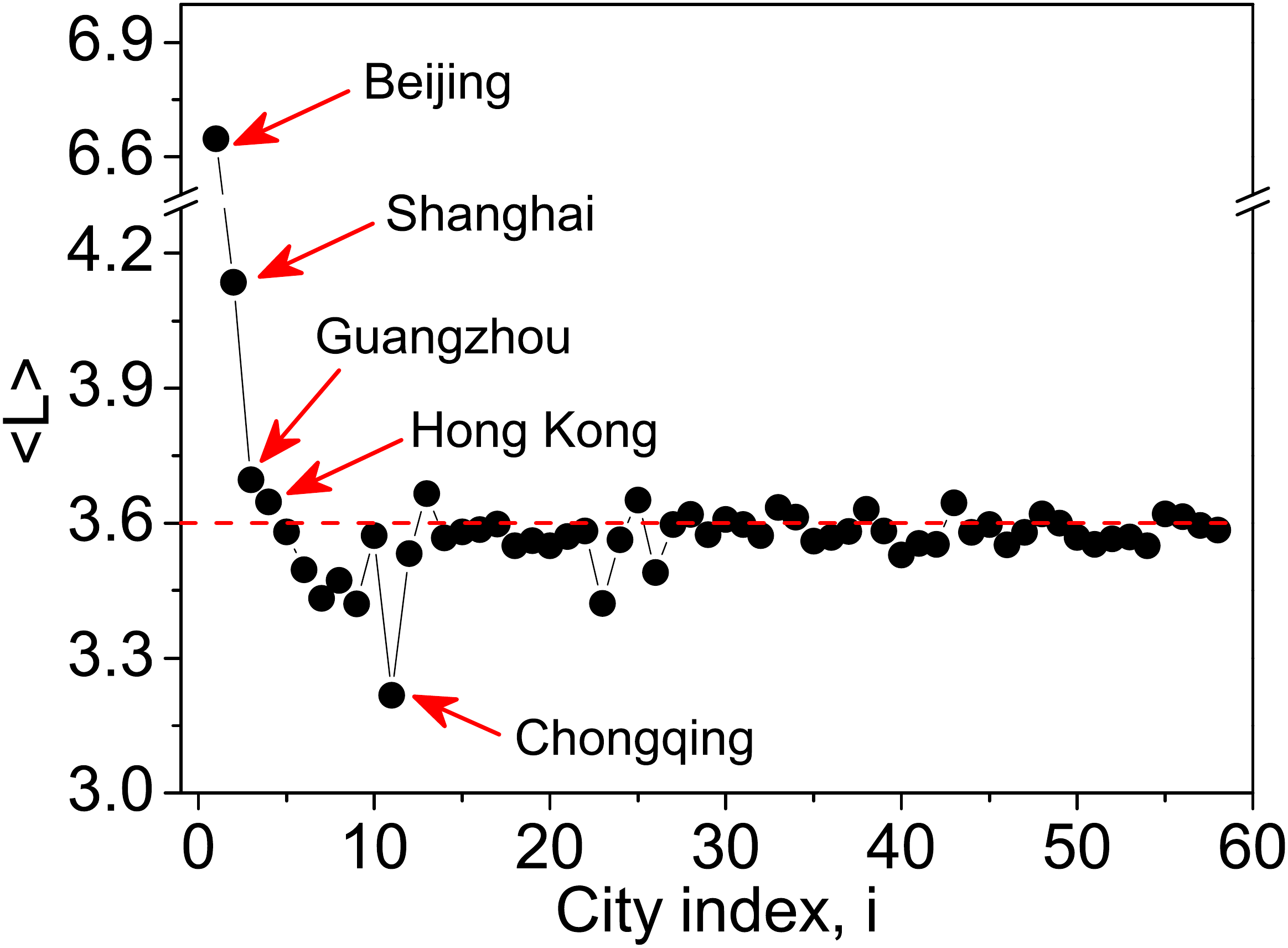}
\caption{{\bf The impact of closing a single node on the averaged route length.} By the results of numerical simulations, the variation of the averaged route length, $\left<L\right>$, with respect to the index, $i$, of the removed city. Red dashed line denotes the value of $\left<L\right>$ for the original network without node removal.}
\label{fig2}
\end{figure}

\emph{Removing a node.} This scenario mimics the closure of a city during a tourism crisis, e.g., the outbreak of an epidemic disease or an terrorist attack. In simulations, this is implemented by removing a city from the network, together with the links associated with it. If a port city is removed, its arrival agents will be distributed to other three port cities in proportion to their arrival volumes. We first investigate how the closure of a city will affect the overall tourist volume of the network, $S=\sum s_i$. As $S=N\left<L\right>$ (with $N$ the total number of agents and $\left<L\right>$ the averaged route length), the value of $S$ is determined solely by $\left<L\right>$. Based on the numerical results, we plot in Fig. \ref{fig2} the variation of $\left<L\right>$ with respect to the index of the closed city, $i$. The results in Fig. \ref{fig2} can be interpreted as follows. (1) The removing of a port city always increases the averaged travel length, therefore increasing the overall tourist volume. Comparing to Guangzhou ($i=3$) and Hong Kong ($i=4$), the averaged travel length is more significantly increased by removing Beijing ($i=1$) or Shanghai ($i=2$). (2) The removing of a non-port city may either increase or decrease the averaged travel length. However, comparing to the port cities, the overall tourist volume is only slightly changed by removing a non-port city. (3) An exception in the non-port cities is Chongqing ($i=11$), where a sharp decrease of the averaged travel length is observed. 

\begin{figure}
\centering
\includegraphics[width=0.9\linewidth]{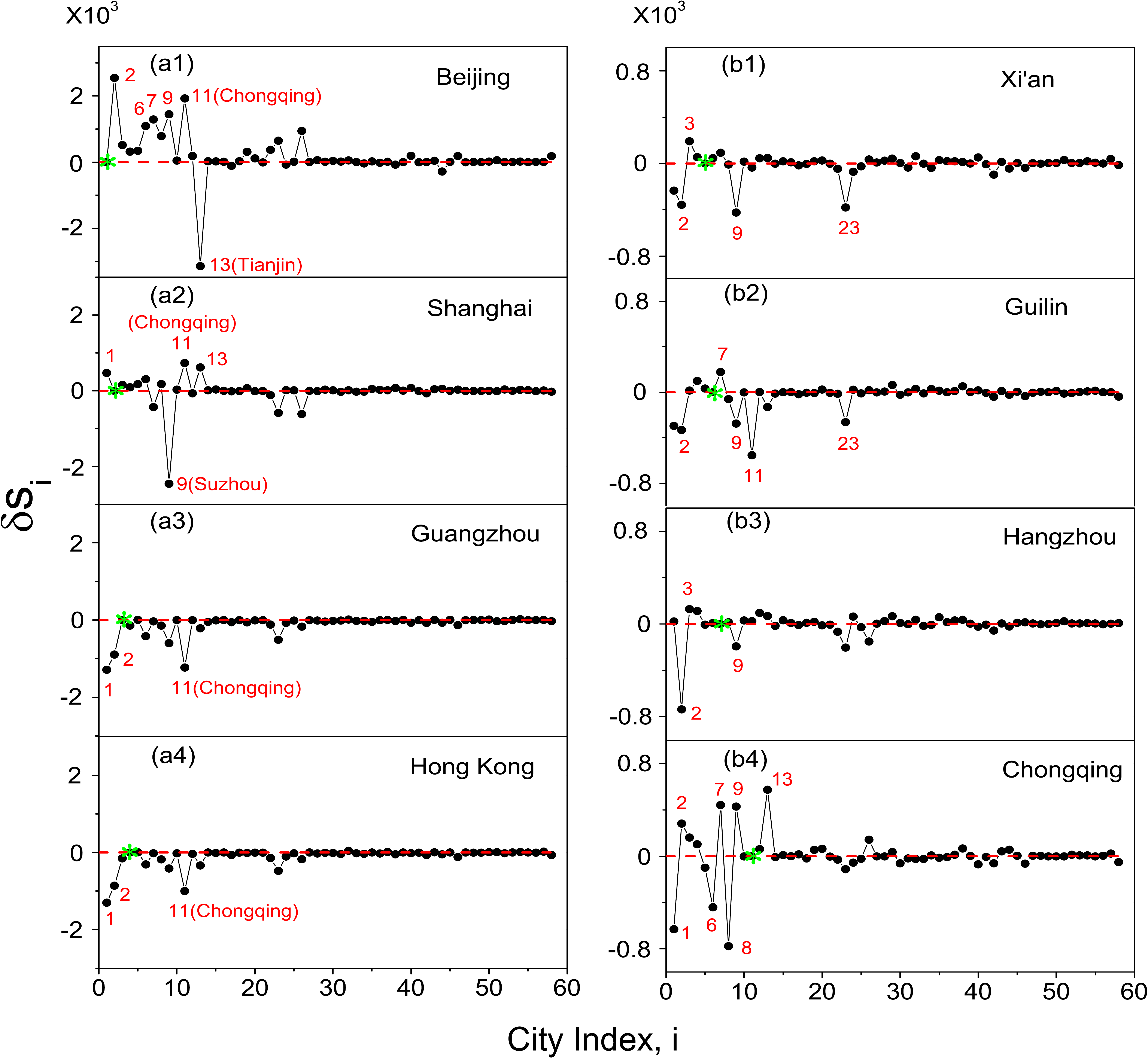}
\caption{{\bf The influence of node removal on city tourist volumes.} Left column: the results for the port cities. Right column: the results for four important non-port cities. $\delta s_i=s'_i-s_i$, with $s'_i$ and $s_i$ the visitation volumes of city $i$ with and without the node removal. The removed city is marked by green asterisk in each panel. Please see Tab. \ref{table1} for the city indices.}
\label{fig3}
\end{figure}

To have more details on the influence of node closure on the tourist distribution, we plot in Fig. \ref{fig3} the change of city visitation volume, $\delta s_i=s'_i-s_i$, when a specific city is removed. Here $s'_i$ denotes the new visitation volume of the $i$th city. The results for removing port cities are plotted in Fig. \ref{fig3}(a1-a4). Figure \ref{fig3}(a1) shows that the removal of Beijing results in a dramatic change of $s_i$ to a few of other cities. By the descending order of $\delta s$, the top $5$ cities that are benefited most from the removal of Beijing are Shanghai ($i=2$), Chongqing ($i=11$), Suzhou ($i=9$), Hangzhou ($i=7$), and Guilin ($i=6$). These five cities are located in different areas on the map [see Fig. \ref{fig1}(a)], signifying the global impact of Beijing on the whole tourism market. Interestingly, it is observed in Fig. \ref{fig3}(a1) that, different from other cities, the visitation volume of Tianjin ($i=13$) is dramatically decreased ($\delta s\approx -3\times 10^3$). The abnormal behavior of Tianjin can be attributed to the spillover effect of the tourists~\cite{YY:2012}. As the adjacent city of Beijing, Tianjin is normally chosen by the tourists as the transition stop. The two cities therefore form a strong collaborative relationship, which leads to the synchronous changes in their visitation volumes. The results for removing Shanghai are plotted in Fig. \ref{fig3}(a2). Similar to the results of Beijing, the removal of Shanghai increases the tourist visitations of a few of other cities too, although with the reduced scales. By the descending order of $\delta s$, the top $3$ cities that are benefited most from the removal of Shanghai are Chongqing, Tianjin, and Beijing. Comparing with the results shown in Figs. \ref{fig3}(a1) and (a2), we see that in the scenario of city closure, the visitation volumes of Beijing and Shanghai are negatively correlated. That is, the closure of one city will increase the visitation of the other city. Figure \ref{fig3}(a2) shows also that the visitation volume of Suzhou is mostly decreased when Shanghai is removed. Again, this phenomenon can be attributed to the spillover effect of the tourists, as Suzhou is in close proximity to Shanghai [see Fig. \ref{fig1}(a)]. It is worth noting that, different from other non-port cities, the visitation volume of Chongqing is significantly increased in both cases, indicating the unique role of Chongqing in the network. 

The results for other two port cities, Guangzhou and Hong Kong, are presented in Figs. \ref{fig3}(a3) and (a4), respectively. Unlike Beijing and Shanghai, the removal of Guangzhou or Hong Kong generates only moderate changes to the visitation volumes of a few other cities. Comparing the results presented in Figs. \ref{fig3}(a3) and (a4), it is interesting to see that the responses of the other cities to the removal of Guangzhou are almost identical to that of Hong Kong. In both cases, the three mostly influenced cities are Beijing, Shanghai, and Chongqing. Particularly, the tourist visitations of Beijing and Shanghai are reduced by approximately the same amount in both cases. The results in Figs. \ref{fig3}(a3) and (a4) suggest that Guangzhou and Hong Kong play a similar role in the network and, different from Beijing and Shanghai, their visitation volumes are positively correlated with each other. The similarity between Guangzhou and Hong Kong could be explained partially by their close proximity, and partially by their remote distance from Shanghai and Beijing [see Fig. \ref{fig1}(a)]. In Figs. \ref{fig3}(a3) and (a4), the responses of Chongqing are distinctly large, implying again the significant role of Chongqing in the network. Different from the cases of Beijing and Shanghai, here we see that the visitation volume of Chongqing is positively correlated with the that of Guangzhou and Hong Kong. That is, the decrease of the visitation volume of Guangzhou (or Hong Kong) leads to the decreased visitation volume of Chongqing.

We move on to evaluate the impact of removing a non-port city on the visitation volumes of other cities. Following the results in Tab. \ref{table1} and Fig. \ref{fig2}, we select Xi'an, Guilin, Hangzhou, and Chongqing as the test cases. The first three cities are chosen for their importance in tourism, i.e., they have the largest visitation volumes among the non-port cities (see Tab. \ref{table1}), while Chongqing is chosen for its unique role in the network [as depicted in Figs. \ref{fig2} and \ref{fig3}(a1-a4)]. The results for removing Xi'an and Guilin are plotted in Figs. \ref{fig3}(b1) and \ref{fig3}(b2), respectively. We see that the removal of Xi'an or Guilin has a modest impact on the visitation volumes of other cities. Figure \ref{fig3}(b3) shows the results for removing Hangzhou. We see that, while the visitation volumes of most of the cities are slightly changed.  the visitation volume of Shanghai is sharply decreased. The strong connection between Shanghai and Hangzhou might be attributed to the spillover effect, as the two cities are adjacent geographically and the tourist flow between them is very large [see Fig. \ref{fig1}(a)]. The results for removing Chongqing are presented in Fig. \ref{fig3}(b4). We see that comparing to other three non-port cities, the removal of Chongqing induces dramatic changes to the visitation volumes of many other cities, especially for those important cities possessing large visitation volumes (of index $1<i<9$ in Tab. \ref{table1}). The significant influence of Chongqing on the whole network might be attributed to its unique geographical location (in the western part of China) and special topological structure (a hub node connecting all four port cities on the network). 

In summary, our numerical studies on node removal show that: (1) the overall performance of the network is more significantly influenced by removing a port city than a non-port city; (2) while the removal of a port city will increase the visitation volumes of many other cities, the visitation volume of its adjacent city could be significantly decreased, due to the spillover effect (Beijing on Tianjin and Shanghai on Suzhou); (3) though as a non-port city, Chongqing is crucially important to the global performance of the network, i.e., the removal of it will induce a large reduction to the overall tourist volume (see Fig. \ref{fig2}), and its visitation volume is strongly coupled to the visitation volumes of many other cities (see Fig. \ref{fig3}).

\emph{Opening a new port city in western China.} To solve the problem of imbalanced tourist distribution (i.e., most of the tourists are concentrated in the east of China), a plausible approach would be upgrading one or a few of cities in the western area, for example, upgrading a non-port city to a port city. As shown in Tab. \ref{table1}, the two most influential cities in western China are Xi'an and Chengdu. These two cities thus are selected in our study as the candidates for upgrading. We also include Chongqing as a candidate due to its unique role in the network. In modeling, when a new port city is opened, we first need to assume two key parameters for it: the percentage of arrival tourists, $p^a_{new}$, and the departure probability of the tourists, $p^d_{new}$. Here, for the purpose of illustration, we set $p^a_{new}=10\%$ and $p^d_{new}=40\%$. The arrival percentages of Beijing, Shanghai, Guangzhou and Hong Kong are adjusted to $35\%$, $25\%$, $15\%$, $15\%$, respectively, while their departure probabilities are kept as unchanged. For simplicity, we assume also that the opening of a new port city does not affect the total number of arrival tourists, i.e., the value of $N$ is kept unchanged.

\begin{figure}
\centering
\includegraphics[width=0.6\linewidth]{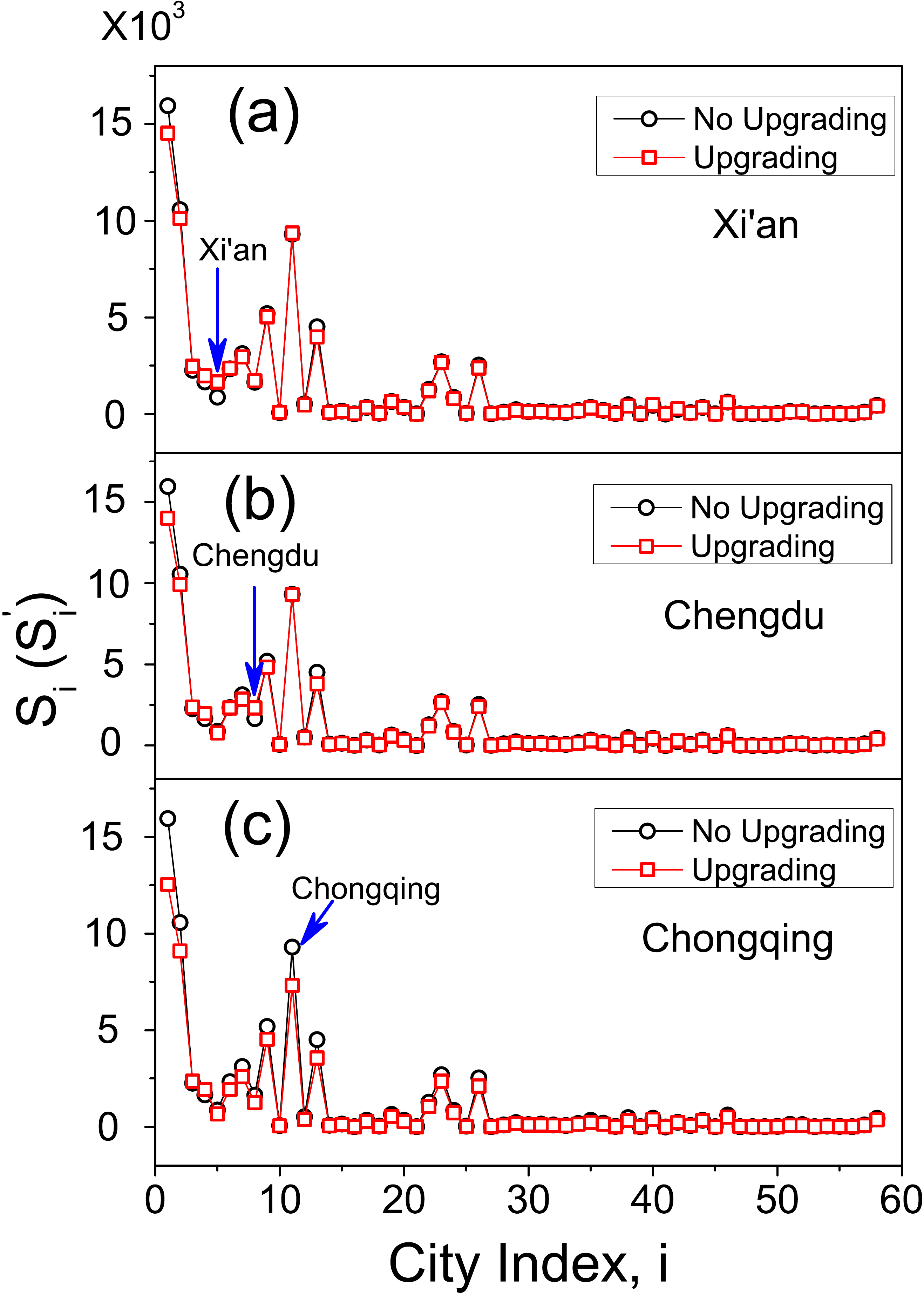}
\caption{{\bf The impact of opening a new port city on the city visitation volumes.} (a) Xi'an. (b) Chengdu. (c) Chongqing. Black circles: the original distribution ($s_i$). Red squares: the distribution when a new port city is opened ($s'_i$).}
\label{fig4}
\end{figure}

We first check the change of the overall tourist volume, $S$, induced by opening a new port city. For the original network, the overall visitation volume is $S=68,302$. When Xi'an is upgraded to a port, the overall visitation volume is increased to $S=65,956$. For Chengdu and Chongqing, the overall visitation volumes are changed to $S=65,104$ and $53,940$, respectively. The numerical results thus suggest that the opening of a new port city in western China has negligible impact on the overall visitation volume. Figure \ref{fig4} shows the changes of the city visitation volumes, $\{s_i\}$, when different cities are upgraded as the port city. We see that, comparing to the results without upgrading, the visitation volumes of the cities are almost unchanged in all cases. To quantify the overall impact of opening a new port city on the tourist distribution, we calculate the change of the geographic concentration index. The geographic concentration index is defined as~\cite{GCI}
\begin{equation}
G=100\times\sqrt{\sum_{i=1}^n(s_i/\left<s\right>)^2},
\end{equation}   
with $n=58$ the number of nodes in the network and $\left<s\right>=S/n$ the averaged city visitation volume. In general, the larger is $G$, the more heterogeneous and imbalanced are the tourists distributed on the network. It is therefore desirable that by opening a new port city in western China the value of $G$ could be significantly decreased, so that more tourists will be attracted from the east to west of China. For the original network, we have $G=36.79$. By upgrading Xi'an, Chengdu and Chongqing as the port city, the value of $G$ is decreased to $33.01$, $33.26$ and $33.27$, respectively. We see that by opening a new port city in western China, the value of $G$ is only decreased slightly, indicating the infeasibility of this approach in balancing the tourist distribution.

In summary, the opening of a new port city in western China affect neither the overall visitation volume, nor the tourist distribution. That is, the problem of imbalanced tourist distribution can not be solved effectively by opening a new port city in western China.

\emph{Increasing city tourism attractiveness.} In tourism development, instead of a global upgrading (due to the limited resources), a common approach adopted in practice is increasing the tourism attractiveness of some specific cities each time, saying, for example, opening a new scenic area in a tourism city. While this approach in general could attract more tourists to the upgraded city, it may either increase or decrease the visitation volumes of other cities, due to the complex collaborative-competitive relationships between the networked cities. Furthermore, for the limited amount of resources, to which city should one invests such that the overall visitation volume of the system is maximally increased? Finally, when a large amount of resources are available, should one invests all the resources into a single city or, alternatively, distributes the resources among a few cities? We next address these questions by numerical simulations.     

\begin{figure}
\centering
\includegraphics[width=0.6\linewidth]{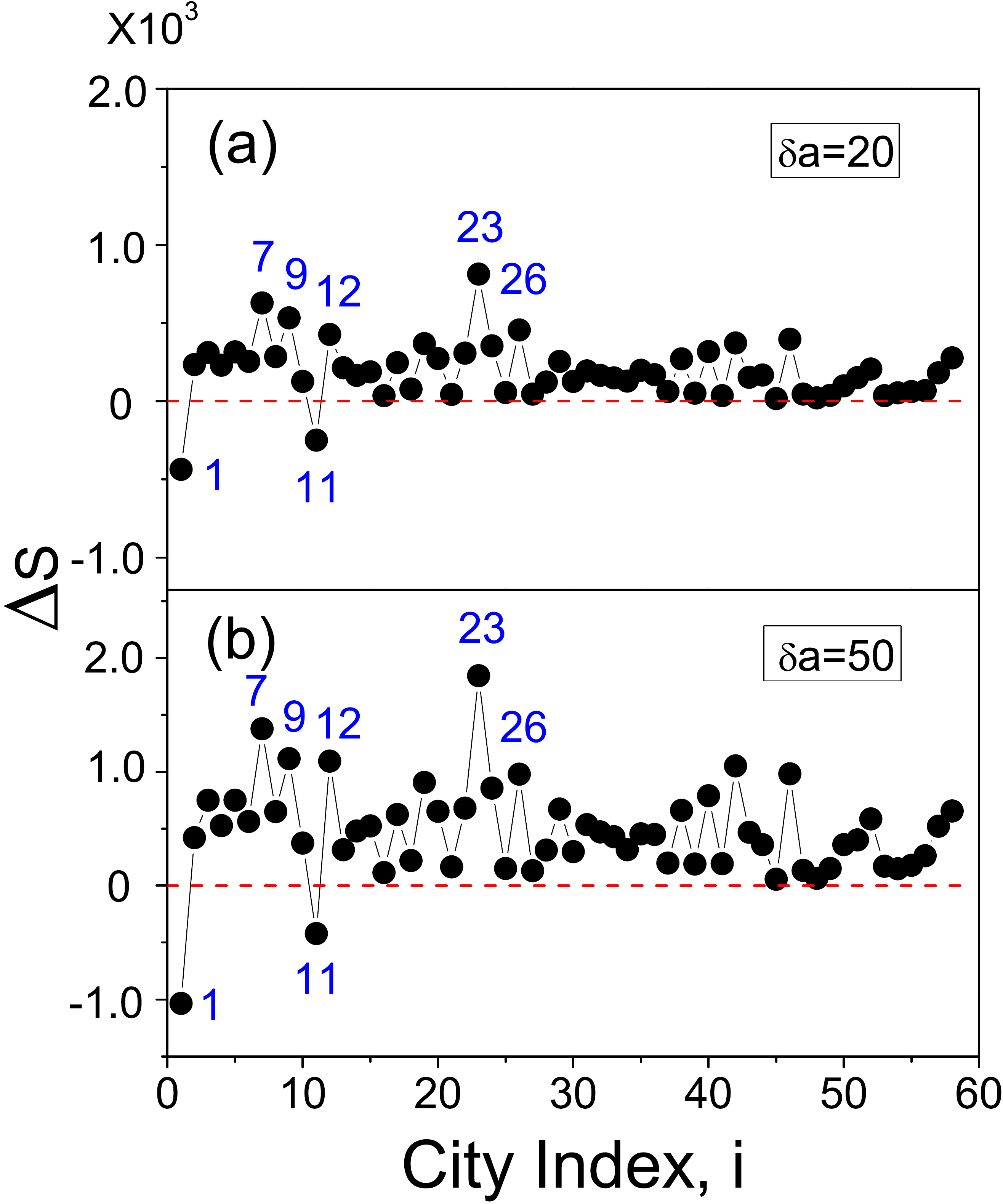}
\caption{{\bf The impact of increasing the tourism attractiveness of a single city on the overall visitation volume.} The increment of tourism attractiveness is fixed as $\delta a=20$ in (a) and $\delta a=50$ in (b). $\Delta S=S'-S$ denotes the change of the overall visitation volume.}
\label{fig5}
\end{figure}

We first evaluate the influence of upgrading a city on the overall tourism performance. Fixing the increment of the tourism attractiveness as $\delta a=20$ [which, according to the calculation of tourism attractiveness (see {\bf Methods}), corresponds to adding $4$ 5A scenic areas in a city], we upgrade the cities individually and calculate the new overall visitation volumes, $S'$. The variation of the overall visitation volume, $\Delta S=S'-S$, with respect to the city index is plotted in Fig. \ref{fig5}(a). We see that, except Beijing ($i=1$) and Chongqing ($i=11$), the upgrading of a city will always increase the overall visitation volume. By the descending order of $\Delta S$, the top $5$ cities are Wuxi ($i=23$), Hangzhou ($i=7$), Suzhou ($i=9$), Nanjing ($i=12$), and Ningbo ($i=26$). As shown in Fig. \ref{fig1}, these $5$ cities are all located in the Yangtze-River-Delta area (the well-developed area in eastern China) and, more importantly, they form a strong subnetwork which attracts a large fraction of the tourists (see Tab. \ref{table1}). This observation seems to suggest that, to increase the overall visitation volume efficiently, the upgrading of a minor city belonging to a strong subnetwork could achieve a better performance. The decreased overall visitation volume at Beijing may be attributed to its higher departure probability ($p^d_1=72.7\%$), which is clearly larger than that of other port cities ($p^d_2=57\%$ for Shanghai, $p^d_3=48.6\%$ for Guangzhou, and $p^d_4=51.1\%$ for Hong Kong). In fact, the impact of upgrading a port city on the overall visitation volume is double-edged. On one hand, the upgrading will attract more tourists to the upgraded city, which tends to increase the overall visitation volume. On the other hand, as the departure probability is keeping unchanged, there will be more tourists departing from the upgraded city, leading to the decrease of the overall visitation volume. If the latter plays the dominant role, the overall visitation volume will be decreased, as in the case of Beijing; otherwise, for the opposite case, the overall visitation volume will be increased, e.g., cities such as Shanghai, Guangzhou, and Hong Kong. The sharp decrease of $\Delta S$ at Chongqing may be attributed to its strong connections to all four port cities. To be specific, when more tourists are attracted to Chongqing, the chance for them to travel to the port cities and then leave China will also be increased. To check the generality of the obtained results, we increase the increment of tourism attractiveness to $\delta a=50$ and plot in Fig. \ref{fig5}(b) again the variation of $\Delta S$ with respect to the city index. We see that the set of most influential cities are not changed by varying $\delta a$, and a sharp decrease of $\Delta S$ is also observed at Beijing and Chongqing.

\begin{figure}
\centering
\includegraphics[width=0.8\linewidth]{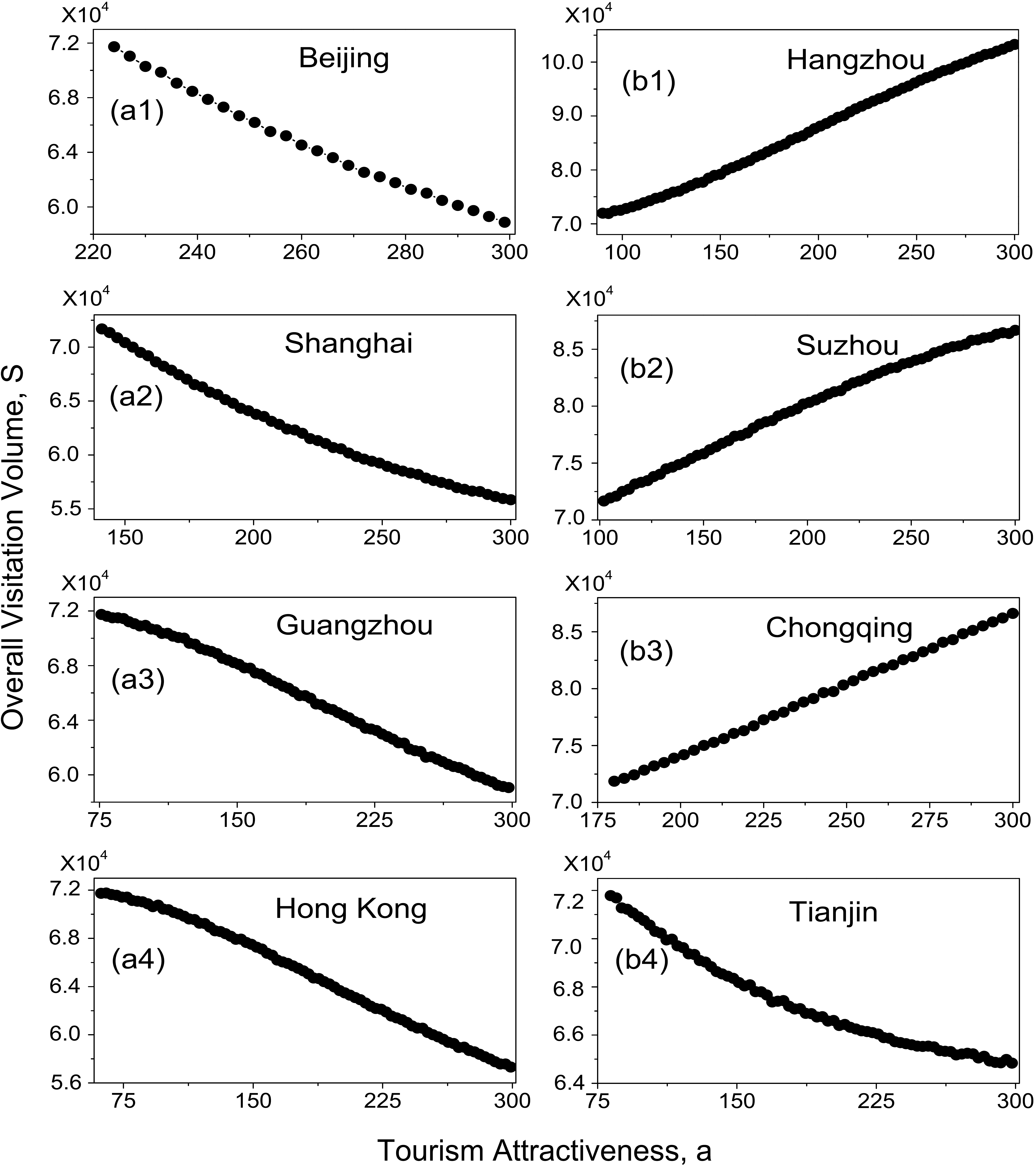}
\caption{{\bf The variation of the overall visitation volume, $S$, with respect to tourism attractiveness, $a$, for the port cities (a1-a4) and several important non-port cities (b1-b4)}. (a1) Beijing. (a2) Shanghai. (a3) Guangzhou. (a4) Hong Kong. (b1) Hangzhou. (b2) Suzhou. (b3) Chongqing. (b4) Tianjin.}
\label{fig6}
\end{figure}

To have a global picture on the influence of tourism attractiveness on overall visitation volume, we investigate the variation of $S$ with respect to $a$ in a wide range, for example, increasing $a$ from its current value to $300$ for each city. The results for the port cities are plotted in Figs. \ref{fig6}(a1-a4). We see that, with the increase of $a$, the value of $S$ is decreased monotonically in each case. This result is expectable, as by increasing the tourism attractiveness more tourists will be attracted to the upgraded port city, which makes the number of tourists departing from it increased. Figures \ref{fig6}(b1-b4) show the variation of $S$ with respect to $a$ for some important non-port cities, including Hangzhou, Suzhou, Chongqing, and Tianjin. Except Tianjin, the value of $S$ is monotonically increasing with the increase of $a$, indicating the positive role of upgrading a non-port city in increasing the overall visitation volume. The abnormal behavior of Tianjin shown in Fig. \ref{fig6}(b4), again, can be attributed to the spillover effect of Beijing, as the visitation volumes of Tianjin and Beijing are positively correlated [see Fig. \ref{fig3}(a1)]. This abnormal behavior, however, is absent when upgrading Suzhou, despite the spillover effect it receives from Shanghai. As depicted in Fig. \ref{fig6}(b2), with the increase of the tourist attractiveness of Suzhou, the overall visitation volume is monotonically increased. The behavior of Suzhou might be attributed to the strong couplings between Suzhou and other cities inside the Yangtze-River-Delta area [see Fig. \ref{fig1}(c)], which reduce the fraction of tourists traveling to Shanghai.    

When the tourism attractiveness of a non-port city is very large, it will be hard to increase the overall visitation volume further by increasing $a$, as the impact of a single node on the whole network has its upper limit. This raises the question of upgrading efficiency, i.e., the increment of overall visitation volume per unit of increase of tourism attractiveness. Conventionally, the upgrading efficiency can be measured by the quantity $S'_{slope}=|\partial S/\partial a|$. Generally, the smaller is the value of $S'_{slope}$, the lower will be the upgrading efficiency. (For the port cities, the value of $S$ is decreased by increasing $a$. In such a case, a smaller value of $S'_{slope}$ represents a slower decrease of $S$.) By scanning $a$ over a wide range, i.e., from its current value to $1\times 10^3$, we calculate the variations of $S$ with respect to $a$ by upgrading different cities (see {\bf Supplementary}), based on which the saturation points (defined as the point where $S'_{slope}=0.1$) can be obtained. The saturation points of some major cities are listed in Tab. \ref{table2}, together with the value of $S$ at the saturation points. We see from Tab. \ref{table2} that the current tourism attractivenesses of the non-port cities are far below their saturation points, indicating their large potentials in tourism development. 

\begin{table}[btp]
\centering
\setlength{\abovecaptionskip}{0pt}%
\setlength{\belowcaptionskip}{10pt}
\caption{The development potentials of some major cities in China. Saturation attractiveness is defined as the point where $S'_{slope}=0.1$ in the variation of the overall visitation volume, $S$, with respect to the tourism attractiveness, $a$. See the text for more details.}
\begin{tabular}{p{2cm}<{\centering} p{3cm}<{\centering} p{3cm}<{\centering} p{3cm}<{\centering} }
\toprule
City & Current attractiveness & Saturation attractiveness  & Saturation overall visitation volume\\
\hline
Xi'an & 66 & 755 & $1.3\times 10^{5}$\\
Guilin & 96 & 845 & $1.5\times 10^{5}$\\
Hangzhou & 90 & 670 & $1.2\times 10^{5}$\\
Chengdu & 80 & 910 & $2.1\times 10^{5}$\\
Suzhou & 102 & 500 & $0.9\times 10^{5}$\\
Kunming & 32 & 950 & $1.8\times 10^{5}$\\
Chongqing & 180 & 900 & $1.1\times 10^{5}$\\
\hline
\end{tabular}
\label{table2}
\end{table}

\begin{figure}
\centering
\includegraphics[width=0.9\linewidth]{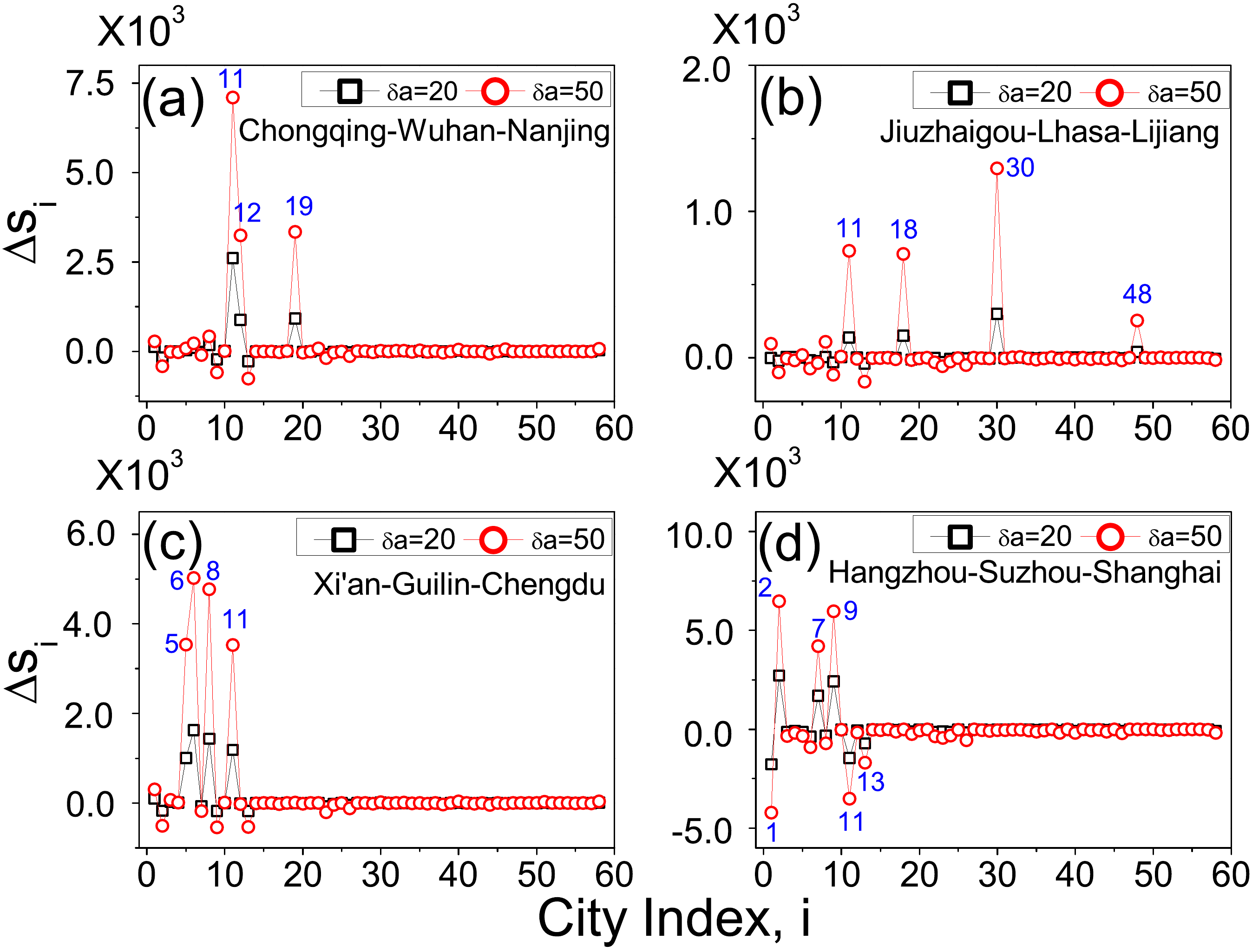}
\caption{{\bf The performance of multiple-city-upgrade}. The changes of city visitation volumes, $\Delta s_i=s'_i-s_i$, by upgrading Chongqing-Wuhan-Nanjing (a), Jiuzhaigou-Lhasa-Lijiang (b), Xi'an-Guilin-Chengdu (c), and Hangzhou-Suzhouo-Shanghai (d). The tourism attractivenesses of the selected cities are increased by the same amount $\delta a$. Black squares: $\delta a=20$. Red circles: $\delta a=50$.}
\label{fig7}
\end{figure}

The saturation point in the variation of overall visitation volume reflects the limit of upgrading a single city in boosting the whole tourism market. When a large amount of upgrading resources are available, $\delta a\gg 1$, instead of upgrading a single city, a better performance might be achieved by distributing the resources among several cities, i.e., the strategy of multiple-city-upgrade. In applying this new strategy, a key question to be addressed is how to select the set of cities giving the best performance. To study, we fix the number of upgraded cities as $3$, and check the performance of the following configurations in improving the overall visitation volume: Chongqing-Wuhan-Nanjing (across China from west to east), Jiuzhaigou-Lhasa-Lijiang (within a small area in western China), Xi'an-Guilin-Chengdu (within a broad area in western China), and Hangzhou-Suzhouo-Shanghai (within the Yangtze-River-Delta area in eastern China). For illustration purpose, we increase the attractiveness of the upgraded cities by the same amount. The changes of the city visitation volumes, $\Delta s_i=s'_i-s_i$ with $s'_i$ the updated city visitation volume, by upgrading Chongqing-Wuhan-Nanjing are plotted in Fig. \ref{fig7}(a). We see that, except the three upgraded cities ($i=11,12,19$), the visitation volumes of the other cities are almost unchanged. As the visitation volumes of the three upgraded cities are clearly increased, the overall visitation volume is still increased by a noticeable amount, as depicted in Tab. \ref{table3}. An interesting finding here is that if the same amount of resources are concentrated in a single city, e.g., Nanjing (which generates the large increment in $S$ among the three cities in terms of single-city-upgrade strategy, as can be seen from Fig. \ref{fig5}), the increment of the overall visitation volume is much smaller to that of multiple-city-upgrade (see Tab. \ref{table3}).

The results for the Jiuzhaigou-Lhasa-Lijiang configuration are plotted in Fig. \ref{fig7}(b). We see that besides the upgraded cities ($i=18,30,48$), the visitation volume of Chongqing (which is not directly upgraded in this case) is also increased significantly. The results for the Xi'an-Guilin-Chengdu configuration are plotted in Fig. \ref{fig7}(c). Similar to the Jiuzhaigou-Lhasa-Lijiang configuration, besides the directly upgraded cities ($i=5,6,8$), the visitation volume of Chongqing is also increased significantly. However, comparing with the results shown in Fig. \ref{fig7}(b), the visitation volumes of the upgraded cities are increased more significantly in Fig. \ref{fig7}(c). As a result of this, the overall visitation volume is significantly increased by the Jiuzhaigou-Lhasa-Lijiang configuration (see Tab. \ref{table3}). Figure \ref{fig7}(d) shows the results for the Hangzhou-Suzhou-Shanghai configuration. Different from the other three configurations, the visitation volumes of Beijing ($i=1$), Chongqing and Tianjin ($i=13$) are clearly decreased. Among the four tested configurations, the Hangzhou-Suzhou-Shanghai configuration generates the smallest increment in the overall visitation volume, as shown in Tab. \ref{table3}. 

\begin{table}[tbp]
\centering
\setlength{\abovecaptionskip}{0pt}%
\setlength{\belowcaptionskip}{10pt}
\caption{The performance of multiple-city-upgrade. Each configuration contains three cities, and the selected cities are upgraded by the same amount of tourism attractiveness, $\delta a$. $S_{original}$, $S_{\delta a=20}$ and $S_{\delta a=50}$ denote, respectively, the overall visitation volume without upgrading, with attractiveness increment $\delta a=20$ and $\delta a=50$.}
\begin{tabular}{p{6cm}<{\centering} p{4cm}<{\centering} p{3cm}<{\centering} p{3cm}<{\centering} }
\toprule
City-configuration & $S_{original}$ & $S_{\delta a=20}$  & $S_{\delta a=50}$\\
\hline
Chongqing-Wuhan-Nanjing & 68,302 & 75,819 & 84,313\\
Jiuzhaigou-Lhasa-Lijiang & 68,302 & 72,211 & 74,163\\
Xi'an-Guilin-Chengdu & 68,302 & 76,451 & 86,871\\
Hangzhou-Suzhou-Shanghai & 68,302 & 72,097 & 72,457\\
\hline
\end{tabular}
\label{table3}
\end{table}

Fig. \ref{fig7} and Tab. \ref{table3} indicate that in applying the multiple-city-upgrade strategy, the target cities should be chosen not only by their local properties (e.g., the tourism attractiveness and visitation volume of the individual cities), but also by their global features (e.g., the geographical location and node degree of the cities). To be specific, network simulations suggest that the more important are the individual cities (with larger attractiveness and visitation volume) and the boarder are the cities distributed on the map, the larger will be the increase of the overall visitation volume. The Jiuzhaigou-Lhasa-Lijiang and Hangzhou-Suzhou-Shanghai configurations take advantages from the former; the Chongqing-Wuhan-Nanjing configuration takes advantage from the latter; while the Xi'an-Guilin-Chengdu configuration takes advantages of both factors. 

Summarizing the influence of increasing city tourism attractiveness on tourism, we have: (1) the increase of tourism attractiveness of a non-port (port) city in general will increase (decrease) the overall visitation volume; (2) the behavior of Tianjin is different from other non-port cities, due to the spillover effect of Beijing; (3) each non-port city has a saturation point in the tourism attractiveness over which the overall visitation volume is hardly increased; (4) when a large amount of upgrading resources are available, the upgrading of several cities could increase the overall visitation volume more efficiently than upgrading just a single city.
 
\section*{Discussions}

In the present study, based on the results of a large-scale survey, we have constructed an agent-based network model for the independent inbound tourism system of China. With this network model, we have investigated numerically the dynamical responses of the tourist flows to external perturbations in certain scenarios of practical significance, including the closure of a city, the opening of a new port city in western China, and the upgrading of city tourism attractiveness. Our main findings are:

\begin{enumerate}

\item The closure of a single city in general will affect the visitation volumes of many other cities in the network, due to the collaborative-competitive relationships between the networked cities. Comparing to a non-port city, the overall visitation volume is more decreased by closing a port city. Whereas the closure of a port city could increase the overall visitation volume in general, there are cities in the network whose visitation volumes are significantly decreased due to the closure, such as Tianjin (when Beijing is closed) and Suzhou (when Shanghai is closed). The abnormal responses of Tianjin and Suzhou are attributed to the spillover effect of the port cities. Among the non-port cities, the behavior of Chongqing is unique in that it is strongly coupled to many important cities in the network, and the closure of Chongqing will generate a significant decrease in the overall visitation volume. 

\item The opening of a new port city in western China will attract more tourists to the western cities, but can not solve fundamentally the problem of imbalanced tourist distribution. As a matter of fact, the opening of a port city in western China will decrease slightly the overall visitation volume. To balance the tourist distribution, a more efficient approach would be increasing the attractiveness of several important city in western China together. Network simulations suggest that with the same amount of upgrading resources, the latter is more efficient in attracting tourists to western China and, in the meantime, increasing the overall visitation volume.

\item The increase of tourism attractiveness of a port (non-port) city in general will decrease (increase) the overall visitation volume. Cities inside the Yangtze-River-Delta area form a strong subnetwork, and behave coherently to external perturbations. More importantly, while individually each city in this subnetwork is not heavily weighted (of larger index in Tab. \ref{table1}), the strong couplings between them render each city with great potential in tourism development, i.e., the increase of the attractiveness of any city in the subnetwork could generate a large increment to the overall visitation volume. An interesting finding is that the upgrade of multiple cities could be more efficient in prompting the China tourism market than upgrading a single city. In choosing the cities to be upgraded, both their individual importance (i.e., tourism attractiveness and visitation volume) and network features (i.e., geographical location and node degree) should be taken into account.

\end{enumerate}

\paragraph*{Implications.} The above findings could be helpful to the development and management of inbound tourism in general, including: 

\begin{enumerate}

\item The results sheds new lights on the dynamic behavior of tourism system. Our numerical studies demonstrate that, due to the collaborative-competitive relationship among the cities, the tourism markets of the tourist cities are not isolated from each other, but are coupled tightly in a dynamic fashion. In particular, it is shown that in some scenarios, e.g., the removal of an important city, the change made on a single city could trigger a large-scale response involving many other cities in the network. Simulation evidences also show that, besides the intrinsic city properties (e.g., the tourism attractiveness), the importance of a city to the whole network is also valued by its structural properties, including the geographical location, node degree, and the set of adjacent cities. These features, which are absent in the model of static tourism networks, manifest the necessary of treating tourism system as a dynamic, complex network.   
	
\item The study provides a quantitative evaluation on the the impacts of external perturbations on the overall tourism performance. With the agent-based network model, we are able to quantify not only the degree of the global response (e.g., the variation of the overall visitation volume), but also the detailed changes at the city level (e.g., the changes of visitation volumes of each city). In addition, we are also able to quantify different relationships between cities, including the spillover effect of Beijing on Tianjin (and also Shanghai on Suzhou), the synchronous behavior of Guangzhou and Hong Kong, and the crucial role of Chongqing.       

\item The study provides a new strategy, namely multiple-city-upgrade, for tourism development. Our simulation results show that neither the opening of a new port city nor the increase of the tourism attractiveness of an individual city in western China could change the imbalanced tourist distribution. To attract more tourists to western China, an efficient approach would be simultaneously upgrading several cities that are deliberately selected according to several factors, including the city importance, connectivity, and geographical location. This finding is of practical significance, as it points out the fact that in developing the overall tourism industry, the tourist cities should be upgraded from the global point of view, instead of their individual interest. 

\end{enumerate} 

\paragraph*{Limitations and future research.} The agent-based network model is constructed based on several key assumptions, which limit the accuracy and applicability of the results obtained. The following lists a few limitations of the proposed model, as well as questions that should be addressed by future studies. 

\begin{enumerate}

\item The number of inbound tourists should not be fixed as constant. For simplicity, we have assumed in network modeling that when a city is closed or upgraded, the total number of agents entered into the system is not changed. In realistic situations, the closure of a city, e.g., due to tourism crisis, will decrease the total number of tourists arriving China significantly. Similarly, increasing the tourism attractiveness of a city will generally attract more inbound tourists. For instance, during the period of the 2018 Summer Olympic Games~\cite{LXY:2012}, Beijing witnessed a large amount of increase in foreign tourists. Besides, the number of inbound tourists also varies with seasons and is affected by the performance of the global economy. Therefore, to make the model more realistic, the number of agents in simulations should be time- and perturbation-dependent.

\item More factors should be included in calculating the city tourism attractiveness, and more details about the tourist travel behavior should be considered. In our model, the attractiveness of a city is determined by only its 4A and 5A scenic areas. In the realistic situation, the city attractiveness is a global tourism indicator incorporating many elements, including economy level, hotels, services, and, of course, tourism attractions. In terms of the tourist travel behavior, we have assumed that the tourists are traveling independently in the network with the probability defined by Eq. (\ref{eq1}). Although the travel routes are made by the tourists independently, there is still the possibility that the tourists interact with each other in an indirect fashion. For instance, a tourist may adjust his/her travel route after reading the travel logs posted by other tourists on websites such as Facebook and Twitter. Finally, in determining the exponent $\gamma$ in Eq. (\ref{eq1}), we have defined it as the average of the optimal values obtained from three different distributions (i.e., the travel lengths, city visitation volumes, and tourist flows). While this definition is meaningful from the point of view of theoretical study, the realistic situation is far more complex than this. Future studies taking into account these issues will be important and necessary.

\item The parameters of the model, either trained from the surveyed results or assumed based on experience, are just for the purpose of demonstration, which should be improved by further empirical studies. In estimating the key parameter $\gamma$ in Eq. (\ref{eq1}), we have adopted simply the average of three optimal values, namely $\gamma_L$, $\gamma_w$ and $\gamma_s$. While these optimal values reflect the collective behavior of the tourists from different aspects, they are essentially related with each other. However, due to the limited surveys and small network size, we are not able to derive such a relation explicitly. Furthermore, besides the analyzed distributions, there could be also other statical quantities that are interested in practical applications. In such a case, the value of $\gamma$ should be redefined according to the specific question that is interested. The similar concern arises also when opening a new port city in western China, in which we have set artificially the parameters, including the fraction of arrival tourists, the arrival percentage and departure probability. Certainly, by changing these parameters, the numerical results, e.g., the variations of the geographic concentration index and the overall visitation volume, will be modified. It is the our hope that by employing more sophisticated analyzing methods and additional empirical data, these parameters could be accurately defined. Nonetheless, our model of agent-based tourism network provides a solid step for the quantitative analysis of the dynamic behavior of complex tourism system. 

\end{enumerate}

\section*{Methods}

\paragraph*{\bf Calculating the city tourism attractiveness.}
The tourism attractiveness of a city is calculated from its tourism attractions as follows. The tourism scenic areas in China are ranked from 1A to 5A by a descending order of aesthetic quality. Among them, the 5A and 4A scenic areas are evaluated based on a national standard, namely the Quality Ranking and Evaluation of Tourism Scenic Areas, produced by the Ministry of Culture and Tourism of China. The scenic areas with ranks from 1A to 3A, however, are evaluated by provincial tourism bureaus on behalf of the national committee. These attractions are mainly targeting domestic tourists. As we are interested in the behaviors of the inbound foreign tourists, we therefore consider only tourism attractions ranked 4A and 5A in calculating the tourism attractiveness. By the time the surveys were conducted, there are totally $117$ 5A scenic areas and $1,849$ 4A scenic areas in China. For the set of cities ($58$ in total) contained in our tourism network model, there are totally $105$ 5A scenic areas and $783$ 4A scenic areas. 

The weights of the scenic areas are obtained by the Delphi method. Specifically, we surveyed all faculty members (6 full, 6 associate, and 6 assistant professors) in the college of the first author, and conducted four rounds of surveys. We adopted the integers for all rounds. In the final results, one 5A scenic area was assigned a value of $5$ and one 4A scenic area was assigned a value of $3$. The full list of the tourism attractiveness of the investigated cities is given in Tab. \ref{table1}. 

\begin{figure}
\centering
\includegraphics[width=0.8\linewidth]{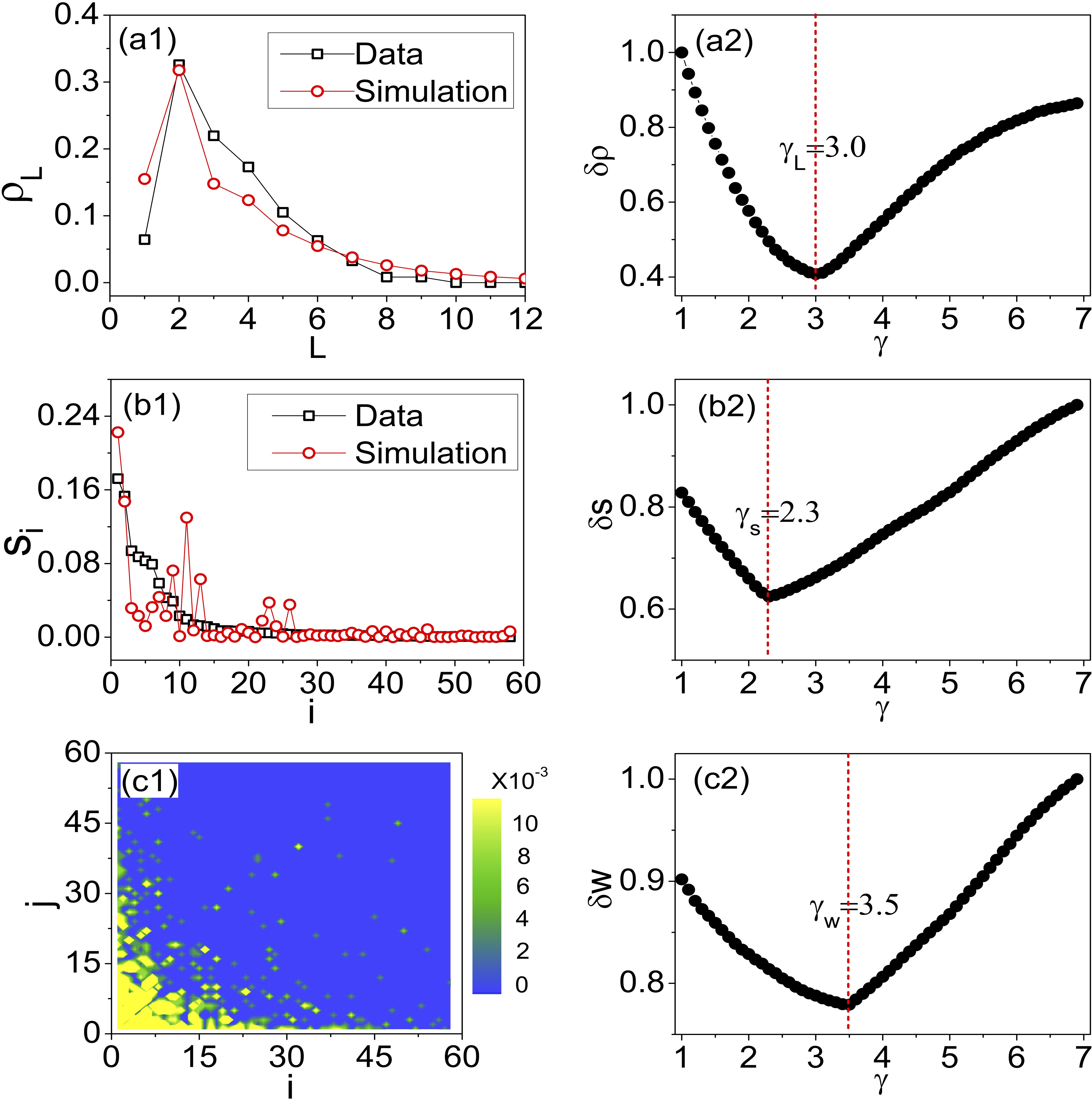}
\caption{{\bf The estimation of parameter $\gamma$}. (a1) The probability distribution of the travel length. Black squares: surveyed results. Red circles: numerical results obtained at $\gamma_L$. (a2) The variation of the normalized difference of travel length, $\delta \rho$, with respect to $\gamma$. $\delta \rho$ is minimized at $\gamma_L\approx 3.0$. (b1) The distribution of city visitation volumes, $\{s_i\}$. Black squares: surveyed results. Red circles: numerical results obtained at $\gamma_s$. (b2) The variation of the normalized difference of city volume, $\delta s$, with respect to $\gamma$. $\delta s$ is minimized at about $\gamma_s\approx 2.3$. (c1) The distribution of the tourist flows, $\{w_{ij}\}$ obtained from surveys. (c2) The variation of the normalized difference of tourist flows, $\delta w$, with respect to $\gamma$. $\delta w$ is minimized at $\gamma_L\approx 3.5$. The parameter $\gamma$ used in Eq. (1) is defined as the average of the three optimal parameters, $\gamma=(\gamma_L+\gamma_s+\gamma_w)\approx 2.9$.}
\label{fig8}
\end{figure}

\paragraph*{\bf Estimating the parameter $\gamma$ based on the surveyed results.} The exponent $\gamma$ in Eq. (\ref{eq1}) is trained by comparing the numerical and surveyed results for three statistical distributions: the route lengths ($\{L_l\}$), the city visitation volumes ($\{s_i\}$), and the tourist flows ($\{w_{ij}\}$). The distribution of the route lengths as obtained from the surveyed results are plotted in Fig. \ref{fig8}(a1), where $\rho_L$ is the fraction of travel routes of length $L$. To find the optimal value $\gamma$ that fits this distribution best, we scan $\gamma$ over the range $(0,7)$ and, by numerical simulations, calculate for each value of $\gamma$ the averaged difference between the the numerical and surveyed results, $\left<\Delta \rho\right>=\sum_{L=1}^{L_{max}}|\rho^s_L-\rho_L|/L_{max}$, with $\rho^s_L$ the fraction of routes of length $L$ obtained by simulations and $L_{max}$ the maximum travel length in the surveys. In Fig. \ref{fig8}(a2), we plot the variation of the normalized average difference, $\delta \rho=\left<\Delta \rho\right>/\left<\Delta \rho\right>_{max}$ ($\left<\Delta \rho\right>_{max}$ is the largest difference in the scanned range), with respect to $\gamma$. We see that $\delta \rho$ reaches its minimum at $\gamma_L\approx 3.0$. We thus choose $\gamma_L= 3.0$ as the optimal parameter trained from the distribution of route length. 

The distribution of the city visitation volumes, $\{s_i\}$, as obtained from the surveys are plotted in Fig. \ref{fig8}(b1). To find the optimal parameter, $\gamma_s$, for this distribution, we calculate the averaged difference between the surveyed and numerical results, $\left<\Delta s\right>=\sum^{n}_{i=1} |s^s_{i}-s_i|/n$, and plot in Fig. \ref{fig8}(b2) the variation of the normalized average difference, $\delta s=\left<\Delta s\right>/\left<\Delta s\right>_{max}$, with respect to $\gamma$. It is seen that $\delta s$ is minimized at about $2.3$. We therefore choose $\gamma_s=2.3$. The distribution of the tourist flows, $\{w_{ij}\}$, as obtained from the surveys are plotted in Fig. \ref{fig8}(c1). Defining the averaged difference as $\left<\Delta w\right>=\sum_{i>j}|w^s_{ij}-w_{ij}|/[(n-1)^2/2]$, we plot in Fig. \ref{fig8}(c2) the variation of the normalized average difference, $\delta w=\left<\Delta w\right>/\left<\Delta w\right>_{max}$, with respect to $\gamma$ based on numerical simulations. We see that $\delta w$ reaches its minimum at $\gamma_w\approx 3.5$. To balance between the three statistical distributions, we take their average, $\gamma=(\gamma_L+\gamma_s+\gamma_w)\approx 2.9$, as the final exponent used in Eq. (\ref{eq1}). 

\section*{Data Availability} 
The data that support the findings of this study are available from the corresponding author upon reasonable request.

{}

\section*{Acknowledgement}

This work was supported by the National Natural Science Foundation of 
China under the Grant Nos. 41428101, 4171135, and 11875132. XGW was also supported by the Fundamental Research Funds for the Central Universities under the Grant No. GK201601001.

\section*{Author contributions}

Devised the research project: JFW and XGW;
Performed numerical simulations: XGW;
Analyzed the results: JFW, XGW, and BP;
Wrote the paper: JFW, XGW, and BP.

\section*{Competing financial interests}

The authors declare no competing financial interest.

\section*{Additional information}

None.

\end{document}